\documentclass[pss]{wiley2sp} 
\usepackage{amsmath}

\newcommand{\kpe}{\mathbf{k}\!\cdot\!\mathbf{p}\,}

\newcommand{\tn}[1]{\textnormal{#1}}
\newcommand{\onlinecite}[1]{\cite{#1}}
\newcommand{\ingan}{\ensuremath{\tn{In}_\tn{x}\tn{Ga}_\tn{1-x}\tn{N}}}
\begin{document}

\bibliographystyle{pss}
\title{Origin of the Broad Lifetime Distribution of Localized Excitons in InGaN/GaN Quantum Dots}

\titlerunning{Origin of the Broad Lifetime Distribution of Localized Excitons in InGaN/GaN QDs}

\author{%
  M.~Winkelnkemper\textsuperscript{\textsf{\bfseries 1,\Ast}},
  M.~Dworzak\textsuperscript{\textsf{\bfseries 1}},
  T.~P.~Bartel\textsuperscript{\textsf{\bfseries 1}},
  A.~Strittmatter\textsuperscript{\textsf{\bfseries 1}},
  A.~Hoffmann\textsuperscript{\textsf{\bfseries 1}}, and
  D.~Bimberg\textsuperscript{\textsf{\bfseries 1}},
  }

\authorrunning{M.~Winkelnkemper et al.}

\mail{e-mail
  \textsf{momme@physik.tu-berlin.de}, Phone
  +49-30-31422074}

\institute{%
  \textsuperscript{1}\,Institut f\"ur Festk\"orperphysik, Technische Universit\"at Berlin, Hardenbergstra{\ss}e 36, 10623 Berlin}

\received{XXXX, revised XXXX, accepted XXXX} 
\published{XXXX} 

\pacs{73.21.La, 78.67.Hc, 78.47.jc}

\abstract{%
We derive an energy-dependent decay-time distribution function from the multi-exponential decay of the ensemble photoluminescence (PL) of \ingan/GaN quantum dots (QDs), which agrees well with recently published single-QD time-resolved PL measurements.  Using eight-band $\kpe$ modelling, we show that the built-in piezo- and pyroelectric fields within the QDs cause a sensitive dependence of the radiative lifetimes on the exact QD geometry and composition. Moreover, the radiative lifetimes also depend heavily on the composition of the direct surrounding of the QDs.  A broad lifetime distribution occurs even for moderate variations of the QD structure. Thus, for unscreened fields a multi-exponential decay of the ensemble PL is generally expected in this material system.\\ (accepted at \emph{Physica Status Solidi (b)}. \copyright 2008 WILEY-VCH)   
 }

%
%

\maketitle   

\section{Introduction}
The \ingan/GaN system has evolved into one of the most important material systems for solid-state light emitters. Applications include green and blue light emitting diodes (LEDs) \cite{nakamura1991,nakamura1998}, laser diodes (LDs) \cite{nakamura1996_2}, and white light emitters \cite{hide1997,sato1996}.  Already at early stage of research the enormous efficiency of \ingan\ optoelectronic devices has been attributed to zero-dimensional localization centers [quantum dots (QDs)] within the \ingan\ layer(s) \cite{chichibu1996,odonnell1999}.  Despite tremendous advances in this research field, the understanding of essential properties of this material system still needs fundamental improvement in order to further advance \ingan/GaN-based optoelectronic devices.  In particular, the influence of the large built-in piezo- and pyroelectric fields in \ingan-based heterostructures on the emission processes is still not understood in detail.  Interestingly, in \ingan\ QDs the quantum-confined Stark effect (QCSE) and the corresponding blue-shift at high charge-carrier concentrations has been reported to be efficiently suppressed \cite{park2007}.  Still, the origin of the \ingan-typical multi-exponential photoluminescence (PL) decay is a matter of active debate \cite{krestnikov2002,bell2004,morel2003,bartel2004}. It has been pointed out by Bartel \textsl{et al.} \cite{bartel2004} that the multi-exponential decay can be consistently explained with broad distribution of single-QD decay times.  In this paper we show that such a broad lifetime distribution---and, consequently, a multi-exponential decay of the ensemble PL---is the effect of the built-in piezo- and pyroelectric fields within the QDs and can be generally expected in this material system.

\section{\label{sec:exp} Sample Preparation and Experimental Results}
The samples investigated in this work have been grown on Si(111) substrate by low-pressure metal-organic chemical vapor deposition (MOCVD) using a horizontal AIX200 RF reactor. An AlAs layer was grown and subsequently converted to AlN as a nucleation surface \cite{strittmatter1999}. In the following step an Al$_{0.05}$Ga$_{0.95}$N/GaN buffer layer was grown at T=1150\,$^{\circ}$C up to a total thickness of 1\,$\mu$m.  The \ingan\ layer was grown at 800\,$^{\circ}$C with a nominal thickness of 2\,nm using trimethylgallium, trimethylindium, trimethylaluminun, and ammonia as precursors. The QDs are formed by alloy fluctuations within the \ingan-layer \cite{seguin2004}.  The growth was finished with a 20\,nm GaN cap layer grown during a heat-up phase to 1100\,$^{\circ}$C.  The PL was excited at 353\,nm by the second harmonic of a mode-locked Ti:sapphire laser.  The temporal width of the laser pulses was 2\,ps at a repetition rate of 80\,MHz.  The measurements were performed in a helium-flow microscope cryostat at a temperature of 4\,K. The luminescence was collected through a microscope objective.  The detection system consisted of two 0.35\,m McPherson monochromators in subtractive mode and an ultrafast photo detector (multi-channel plate) providing a spectral resolution of about $0.6$\,meV and a temporal resolution better than $30$\,ps. The single-QD measurements are described in detail elsewhere \cite{bartel2004}. The PL spectrum of the entire QD ensemble has its maximum at 3.05\,eV and a full width at half maximum of $\approx75$\,meV (inset in Fig.~\ref{intensity}). Sharp emission lines of discrete QD states were observed from $2.8$ to $3.2$\,eV proving the QD origin of the entire emission \cite{bartel2004,seguin2004}. 
\begin{figure}
\includegraphics[width=\columnwidth,clip]{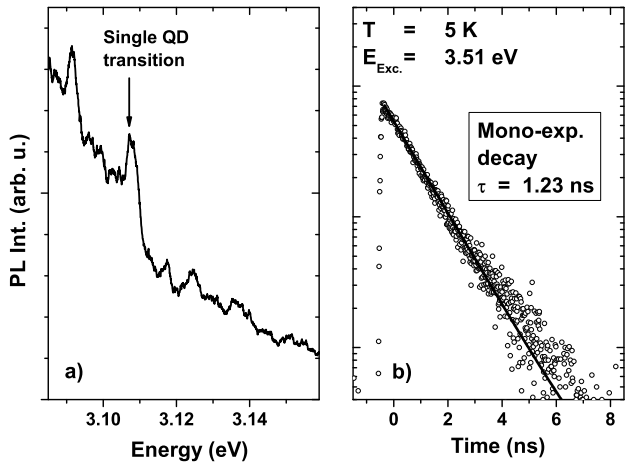}
\caption{\label{singledot} $\mu$-PL spectra of a single \ingan/GaN QD. \emph{(a)} Time-integrated spectrum. \emph{(b)} Time-resolved PL of the emission line indicated by the black arrow in (a). The decay is well described mono-exponentially (solid black line). Ten different single-QD transients have been investigated by Bartel \textsl{et. al} \cite{bartel2004}, all of which showed a mono-exponential decay.} 
\end{figure}

Photoluminescence transients measure on emission lines from single QDs mono-exponential (Fig.~\ref{singledot}) \cite{bartel2004}. The ensemble-PL decay, in contrast, is multi-exponential for all detection energies. A similar behaviour---multi-exponential decay of the macro PL and mono-exponential decay in single-QD measurements---has also been reported by Robinson \textsl{et al.} \cite{robinson2003}. Rice \textsl{et al.} \cite{rice2005} have even convincingly demonstrated that they observe the decay of single excitons and biexcitons in their experiments on single \ingan\ QDs.  Moreover, the shapes of the ensemble-PL transients do not depend on the  excitation density (Fig.~\ref{intensity}), indicating that the multi-exponential decay can be attributed to a broad distribution of decay times \cite{bartel2004} rather than dynamical screening effects \cite{reale2003,gotoh2003}.   
\begin{figure}
\includegraphics[width=\columnwidth,clip]{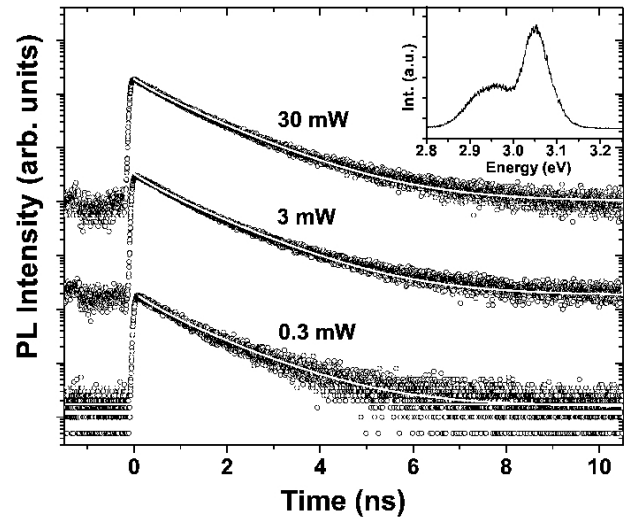}
\caption{\label{intensity} Excitation-density dependence of the TRPL signal at fixed detection energy ($3.05$\,eV; black arrow in the inset). A stretched-exponential function with identical parameters describes all transients (white lines). The inset shows the time-integrated PL of the sample.} 
\end{figure}

\section{\label{sec:f}Distribution of Excitonic Lifetimes}

All QDs with the same excitonic transition energy $E$ can be considered a subensemble within the entire QD ensemble.  The distribution of decay times $\tau$ within each subensemble shall be described by the distribution function $f_E (\tau)$.  This distribution function can be approximated from TRPL measurements of the subensembles: To eliminate the noise of the experimental data, the multi-exponential subensemble PL decays are approximated by the Kohlrausch (or stretched-exponential) function \cite{kohlrausch1854} (solid white lines in Fig.~\ref{intensity})
\begin{equation}\label{stretched_exp}
    I_E(t)=I_{E,0} \exp[-(t/\tau_E^*)^{\beta_E}]\quad.
\end{equation}
Here, $\tau_E^{*}$ and $\beta_E$ are the time and stretching parameters for the given energy $E$. $\tau_E^{*}$ is identical to the decay time of the system $\tau_E$ only if $\beta_E=1$. It has no direct physical meaning for $\beta_E\neq1$.  The decay of the PL signal at eleven different equidistant detection energies between 2.82\,eV and 3.18\,eV has been analyzed. The shapes of the transients depend on the detection energy, resulting in varying fit parameters for different QD subensembles ($0.35$\,ns$\leq \tau_E^* \leq 0.85$\,ns; $0.55\leq \beta_E \leq 0.84$).

The PL decay of each QD subensemble can also be expressed as an integral over the (exponential) single-QD PL decays of all QD which form the subensemble.  Thus, using the Kohlrausch function to describe the decay, we can expand Eq.~\ref{stretched_exp} to

\begin{equation}\label{laplace}
   I_E(t) \propto \exp[-(t/\tau_E^*)^{\beta_E}] \propto \int_0^\infty f_E(\tau)\exp(-t/\tau) d\tau\quad.
\end{equation}

Mathematically, the subensemble decay $I_E(t)$ is, hence, the Laplace transform of the lifetime distribution function $f_E(\tau)$ within the subensemble, which can, in turn, be obtained by an inverse Laplace transformation of $I_E(t)$ \cite{footnote_wolfram,footnote_driel}.  
\begin{figure}
\includegraphics[width=\columnwidth,clip]{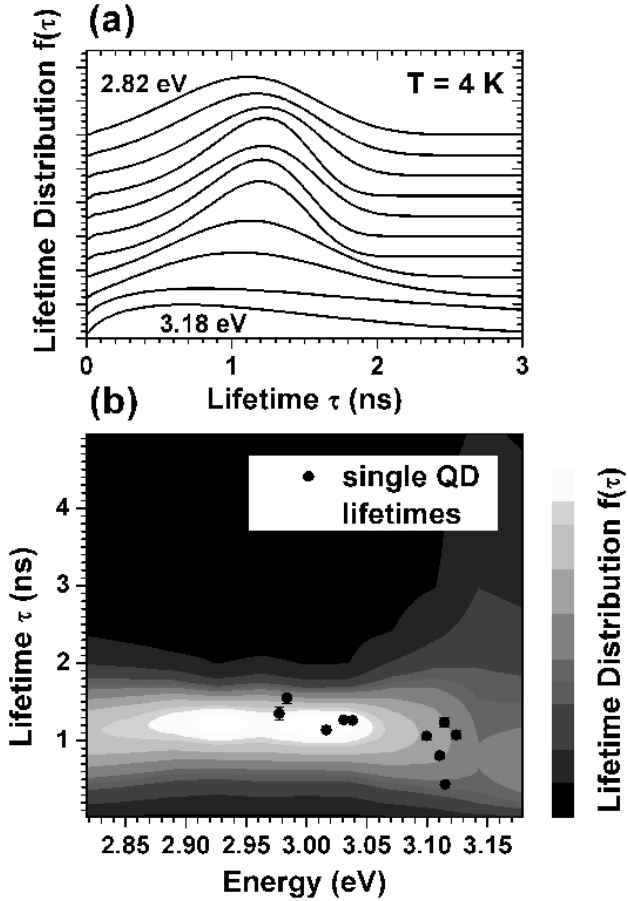}
\caption{\label{distribution} (a) Lifetime distribution functions $f_E(\tau)$, obtained by inverse Laplace transformation of the multi-exponential PL decay of different InGaN-QD subensembles (different detection energies $E$). The integral of each $f_E(\tau)$ is normalized to 1. (b) $f(E,\tau)$ as gray-scale plot together with the decay times of single InGaN QDs reported in Ref.~\onlinecite{bartel2004}. The single-dot lifetimes agree well with the obtained distribution function.}
\end{figure}

The results of the inverse Laplace transformation are shown in Fig.~\ref{distribution}.  A broad distribution of lifetimes for all transition energies is found, in good agreement with the single-QD TRPL results from Ref.~\onlinecite{bartel2004} [black symbols in Fig.~\ref{distribution}(b)], which also show an appreciable scatter for identical transition energies. All single-QD time constants fall in the range covered by the lifetime distribution function.

For GaN/AlN QDs a drastic increase (several orders of magnitude) of the excitonic lifetimes for lower transition energies has been reported \cite{bretagnon2006}, which is caused by the increasing QCSE  for increasing QD height.  Such an effect can not be observed in the lifetime distribution function in Fig.~\ref{distribution}: The maximum of the distribution function is  constant at $\approx1.2$\,ns up to transition energies of $\approx3.05$\,eV; for higher transition energies it shifts slightly to shorter lifetimes, accompanied by a broadening of the distribution. As will be alluded to in Sec.~\ref{sec:kp}, the broadening of the distribution at higher transition energies can be explained with the delocalization of charge carriers that are only weakly bound to shallower localization centers.

\section{\label{sec:kp}Quantum Dot Structure and Radiative Lifetimes}

The broad lifetime distribution within the InGaN-QD ensemble can be understood in terms of varying electron-hole wave-function overlap in different QDs.  We will show here that the variation of the overlap is caused by differences in the built-in piezo- and pyroelectric fields within each localization center.  

We have calculated the radiative lifetimes of different InGaN QDs using a three-dimensional eight-band $\mathrm{\textbf{k} \cdot \textbf{p}}$ model.  The model accounts for strain effects, piezoelectric and pyroelectric polarization, spin-orbit and crystal-field splitting, and coupling between the valence bands (VBs) and the conduction band (CB).  Excitonic corrections have been included using a self-consistent Hartree (mean field) scheme.  A detailed description of the method can be found elsewhere \cite{footnote_method}. The radiative lifetimes $\tau_\tn{rad}$ of the confined excitons have been calculated by \cite{footnote_dexter} 
\begin{equation}\label{eq:tau}
\tau_\tn{rad} =\frac{
2\pi\varepsilon_0m_0c_0^3\hbar^2
}{
ne^2E_{\tn{ex}}^2f_\tn{eff}}
\quad.
\end{equation}
Here, $\varepsilon_0$ is the permittivity of free space, $m_0$ the free-electron mass, $c_0$ the vacuum speed of light, $\hbar$ the reduced Planck constant, and $E_\tn{ex}$ the transition energy of the exciton. $n$ is the refractive index of the matrix material (GaN), which can be described as a function of the emission wavelength by a Sellmeier-type law \cite{footnote_sellmeier}.  The effective oscillator strength $f_\tn{eff}$ has been calculated by integrating the (anisotropic) oscillator strength $f_\mathbf{e}$ over the unit sphere:  

\begin{equation}\label{kp:f}
f_\tn{eff}=\frac{1}{4\pi}\int_{\partial O}f_\mathbf{e}\tn{d}\mathbf{e}\quad;\quad
f_\mathbf{e}=\frac{2\hbar^2}{m_0E_\tn{ex}}\left|\langle\Psi_\tn{e}|\mathbf{e}\cdot\mathbf{\hat{p}}|\Psi_\tn{h}\rangle\right|^2\quad.
\end{equation}
Here, $|\Psi_\tn{j}\rangle$ are the $\kpe$ electron and hole wave functions, which consists of a sum over all $\kpe$ basis states $|i\rangle$ (Bloch functions) multiplied with the respective envelope functions $\varphi_\tn{i}$: $|\Psi_\tn{j}\rangle=\sum_\tn{i=1}^{8}\varphi_\tn{i}|i\rangle$. When calculating the matrix elements contribution arising from the Bloch parts and the envelope-function parts of the wave functions have been accounted for. $\mathbf{e}$ is a unit vector indicating the polarization of the light.

In agreement with our recent publications \cite{winkelnkemper2007,winkelnkemper2006}, the QDs have been modeled as ellipsoids with height $h$ and lateral diameter $d$. They are embedded in a $2$\,nm thick InGaN quantum well (QW) with indium concentration $x_\tn{w}$.  The indium concentration within the QDs increases linearly from the indium fraction of the surrounding QW ($x_\tn{w}$) to the maximum indium concentration $x_\tn{c}$ at the QD center.  Starting with a QD with a height of $h=2$\,nm, a lateral diameter of $d=5.2$\,nm, and indium concentration of $x_\tn{c}=0.5$, the influences of three different structural parameters on the radiative excitonic lifetimes have been investigated: The QD height (with $h$ varying between $1.2$ and $2.8$\,nm), the lateral diameter (with $d$ varying between $2.8$ and $7.6$\,nm), and the indium concentration within the QD ($x_\tn{c}$ between $0.3$ and $0.6$).  All three series have been calculated with the QDs embedded in an $\tn{In}_{0.1}\tn{Ga}_{0.9}\tn{N}$ QW, in an $\tn{In}_{0.05}\tn{Ga}_{0.95}\tn{N}$ QW, and directly in the GaN matrix, without a QW. In the latter case QDs with slightly higher In concentration ($x_\tn{c}=0.7$) or slightly larger diameter ($d=8.8$\,nm), respectively, have been included in order to cover the experimentally observed energy range. Note that, due to the concentration gradient inside the QDs, the average indium concentration within the QDs $x_\tn{avg}$ is much lower than the maximum indium concentration of $x_\tn{c}=0.3$-$0.6$ ($0.7$) at the QD center. A maximum indium concentration of $x_\tn{c}=0.3$ corresponds to an average concentration of $x_\tn{avg}\approx 0.08$-$0.15$ depending on the indium concentration in the QW ($x_\tn{w}=0.0$-$0.1$). An $x_\tn{c}$ of $0.6$ ($0.7$) corresponds to $x_\tn{avg}\approx 0.15$-$0.23$ ($0.18$-$0.25$).  The magnitude of the deviation $\Delta x=x_\tn{avg}-x_\tn{w}\approx 0.08$-$0.15$ is in good agreement with typical values for alloy fluctuations in InGaN QWs \cite{bartel2007}.

\begin{figure}
\includegraphics[width=\columnwidth,clip]{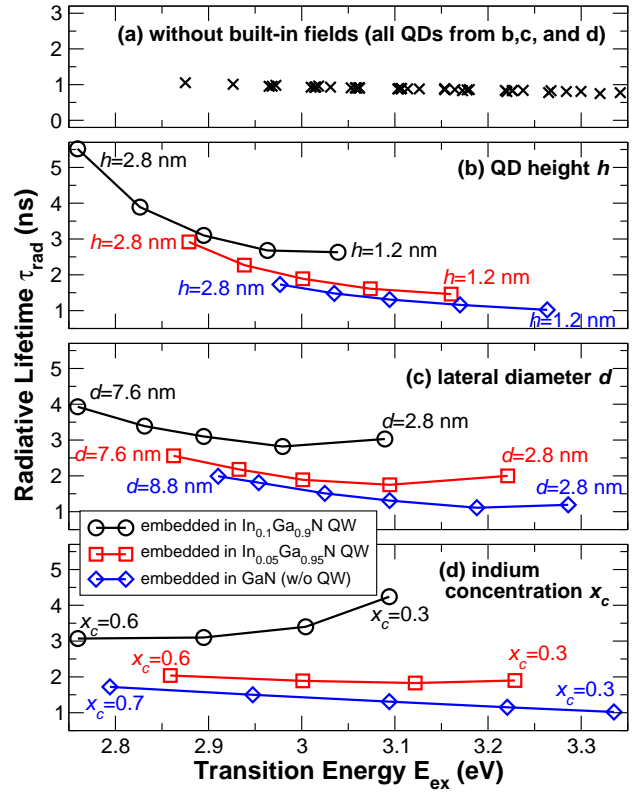}
\caption{\label{kpresults} Radiative excitonic lifetimes vs.\ transition energies calculated with eight-band $\mathrm{\textbf{k} \cdot \textbf{p}}$ theory: (a) All model QDs considered in this paper (see text), neglecting the built-in piezo- and pyroelectric fields. (b-d) As a function of the QD height, diameter, and composition, including piezo- and pyroelectric effects.}
\end{figure}

Neglecting the built-in piezo- and pyroelectric fields [Fig.~3(a)] all QDs show similar excitonic lifetimes around $0.8$-$1.0$\,ns.  The radiative lifetimes are insensitive to the exact QD shape and composition. Such a lifetime distribution would lead to mono-exponential PL decay for all transition energies.  Indeed, such a behaviour has been observed by S\'{e}n\`{e}s \textsl{et al.} \cite{senes2007} for \ingan-QDs grown by molecular beam epitaxy (MBE), were field effects are reported to be negligible \cite{senes2007}.  

For the MOCVD-grown QDs investigated in this work a strong deviation from the mono-exponential decay has been observed for all transition energies. This deviation can be explained if the built-in electrostatic fields are included in the calculations [Fig.~3(b-d)]. The radiative excitonic lifetimes scatter across a wide range between $1.0$ and $5.5$\,ns.  Pronounced dependencies on all structural parameters can be observed.  Interestingly, the radiative lifetimes significantly decrease, when the indium concentration of the surrounding QW is decreased from $10$\,\% to $5$\,\%, or even to $0$\,\%.  Then, the matrix exerts a increasing compressive strain to the $z$-axis ([0001]-axis) of the QD which reduces the built-in field from $\approx 2.5-4.3$\,MV/cm if embedded in a $10$\,\% QW down to $\approx 1.7-3.1$\,MV/cm ($\approx 1.0-2.6$\,MV/cm) in a $5$\,\% ($0$\,\%) QW. The separation of the centers of mass of the electron and hole wave functions decreases from $\approx 0.9-1.6$\,nm to $\approx 0.7-1.2$\,nm ($\approx 0.4-0.8$\,nm). Thus, not only the exact geometry and chemical composition of the QDs themselves, but also the properties of the direct surrounding area of the QDs are decisive for the radiative lifetimes.  A large variety of different time constants can be found even for identical transition energies.  For instance, among all considered model QDs, six different QDs emit at $\mathrm{\approx 3.1~eV}$ (Fig.~3).  Although all six QDs have nearly the same transition energy, their radiative lifetimes scatter appreciably between $1.3$\,ns and $4.2$\,ns.  Thus, luminescence at a certain detection energy originates from a subensemble of QDs, all of which have the same transition energy, but significantly different excitonic lifetimes.

The calculated lifetimes describes the experimental scatter well qualitatively, but quantitatively they are generally larger than the measured ones.  On the one hand the uncertainties of the material parameters of \ingan\ provide a reasonable explanation for this systematic deviation.  
On the other hand, Narvaez \textsl{et al.} \cite{narvaez2005} have shown for InGaAs/GaAs QDs that the lifetimes of charged excitons (positive or negative trion) are shorter than that of the exciton by a factor of $\approx0.5$. Therefore, another possible explanation is that the PL originates from the decay of charged excitons rather than neutral ones. 

\section{Conclusion}
We have extracted the photon-energy-dependent decay-time distribution function from the multi-exponential decay of the ensemble PL of \ingan/GaN QDs, which agrees well with recently publish single-QD decay times.  We have calculated the radiative lifetimes of localized excitons in \ingan/GaN QDs and shown that the built-in piezo- and pyroelectric fields within the QDs are the origin of the broad lifetime distribution: They cause a sensitive dependence of the radiative lifetimes on the QD geometry and composition. The lifetimes are also very sensitive to the chemical composition in the direct vicinity of the dots.   Therefore, a broad distribution of excitonic lifetimes and, consequently, a multi-exponential decay of the ensemble PL is generally expected in this material system in the case of unscreened fields.  A mono-exponential decay of the ensemble luminescence, on the other hand, indicates vanishing field effects. 

\begin{acknowledgement}
The authors are indebted to R.~Seguin, A.~Schliwa, T.~Stempel~Pereira, L.~Rei{\ss}mann, and S.~Rodt for fruitful discussions and critical reading of the manuscript.  This work was supported by the Deutsche Forschungsgemeinschaft in the framework of Sfb 787 and by the SANDiE Network of Excellence of the European Commission, contract number NMP4-CT-2004-500101. Parts of the calculations were performed on an IBM p690 supercomputer at HLRN Berlin/Hannover with project No.~bep00014.
\end{acknowledgement}

\providecommand{\WileyBibTextsc}{}
\let\textsc\WileyBibTextsc
\providecommand{\othercit}{}
\providecommand{\jr}[1]{#1}
\providecommand{\etal}{~et~al.}

\end{document}